\documentclass[11pt]{article}

\usepackage{abstract}
\usepackage{graphicx}
\usepackage{authblk}

\usepackage{natbib}

\usepackage{amsmath}

\usepackage{setspace}

\usepackage{multirow}
\usepackage{booktabs}

\usepackage{subfigure}

\usepackage{url}
\usepackage{hyperref}
\usepackage{amssymb}
\usepackage{amsmath}
\usepackage{ifthen}
\usepackage{caption}
\usepackage{amsthm}
\usepackage{amstext}

\usepackage{tikz}
\usetikzlibrary{shapes,positioning}
\usepackage{bigints}

\newcommand{\C}{{\phi}}
\newcommand{\SRT}{{\Psi}}
\newcommand{\RT}{{\psi}}
\newcommand{\nt}{{n}}
\newcommand{\nc}{{K}}
\newcommand{\HE}{{h}}
\newcommand{\TR}{{i}}

\newcommand{\TRI}[1]{\TR{}_#1}

\newcommand{\NR}{\mathrm{R}}

\newcommand{\G}{\mathit{G}}

\newcommand{\fBD}{f_{\textsc{g}}}

\newcommand{\fc}{\rho_{\textsc{g}}}

\newcommand{\uh}{\rho_h}

\newcommand{\fM}{f_h}

\newcommand{\rd}{^{(R)}}

\newcommand{\RA}{{\bar{i}}}
\newcommand{\RAI}[1]{\RA{}_#1}

\newcommand{\AR}{{\hat{i}}}
\newcommand{\ARI}[1]{\AR{}_#1}

\newcommand{\txa}{a}
\newcommand{\txb}{b}
\newcommand{\txc}{c}
\newcommand{\txd}{d}
\newcommand{\txtr}[1]{\textsc{\small{#1}}}
\newcommand{\txsb}[1]{\textsc{\tiny{#1}}}

\newcommand{\txA}{\txtr{a}}
\newcommand{\txB}{\txtr{b}}

\newcommand{\txAB}{\txA\txB}

\def\D{\mathrm{d}} 

\newcommand{\bydef}{\equiv}

\begin{document}
\title{Calibrated birth-death phylogenetic time-tree priors for Bayesian inference}

\author[1]{Joseph Heled\thanks{jheled@gmail.com}}
\author[1,2]{A.~J.~Drummond\thanks{alexei@cs.auckland.ac.nz}}
\affil[1]{Department of Computer Science, The University of Auckland, Auckland, New Zealand}
\affil[2]{Allan Wilson Centre for Molecular Ecology and Evolution, New Zealand}

\maketitle 
\begin{onecolabstract}

  Here we introduce a general class of multiple calibration birth-death tree
  priors for use in Bayesian phylogenetic inference. All tree priors in this
  class separate ancestral node heights into a set of ``calibrated nodes'' and
  ``uncalibrated nodes'' such that the marginal distribution of the calibrated
  nodes is user-specified whereas the density ratio of the birth-death prior is
  retained for trees with equal values for the calibrated nodes.

  We describe two formulations, one in which the calibration information informs
  the prior on ranked tree topologies, through the (conditional) prior, and the
  other which factorizes the prior on divergence times and ranked topologies,
  thus allowing uniform, or any arbitrary prior distribution on ranked
  topologies.  While the first of these formulations has some attractive
  properties the algorithm we present for computing its prior density is
  computationally intensive. On the other hand, the second formulation is always
  computationally efficient.  We demonstrate the utility of the new class of
  multiple-calibration tree priors using both small simulations and a
  real-world analysis and compare the results to existing schemes.

  The two new calibrated tree priors described in this paper offer greater
  flexibility and control of prior specification in calibrated time-tree
  inference and divergence time dating, and will remove the need for indirect
  approaches to the assessment of the combined effect of calibration densities
  and tree process priors in Bayesian phylogenetic inference.

\end{onecolabstract}

\newpage
\section{Introduction}
Divergence time dating and phylogenetic inference are related concerns. Recent
advances in Bayesian phylogenetic inference
\citep{Rannala:1996uq,Yang:1997uq,huelsenbeck2001mbi,BEAST} have culminated in
the field of relaxed phylogenetic inference, in which both divergence times and
phylogenetic relationships are simultaneously estimated
\citep{drummond2006relaxed}. This estimation is aided by relaxed molecular
clocks
\citep{Thorne1998,kishino:2001:pe,Thorne:2002ff,drummond2006relaxed,Rannala2007}
which reconcile non-clock-like evolution with an underlying time-tree in which
common ancestors are placed on an axis of time. In order to produce results on
an absolute time scale it is necessary to either provide information on the rate
of molecular evolution or alternatively calibrate a subset of internal nodes
with a calibration density
\citep{Thorne1998,yang2006bayesian,drummond2006relaxed}. Either way, In a Bayesian
setting, a {\it tree process prior} must also be placed on all the uncalibrated
divergence times. Arguably the simplest tree process priors are the
one-parameter Yule model \citep{Yule1924} and the two-parameter birth-death
model \citep{Nee:1994fk,gernhard2008conditioned}. The latter has been suggested
as an appropriate null model for species diversification \cite{Nee1994b} and has
been extended to include additional parameters to model various types of
incomplete sampling \citep{Yang:1997uq,Stadler:2009fk,Hohna:2011kx}.

In a Bayesian setting, combining a calibration density (on one or more
divergences) with a tree process prior into a single calibrated tree prior for
divergence time estimation possesses a number of subtleties worthy of note,
which we cover under the following headings.

\subsection{\it Fossil bounds on a single divergence}

Consider the simplest type of calibration to admit uncertainty: the placement of
an upper and a lower limit on the age of a single monophyletic calibrated divergence ($h_{C}$) in
the tree:

\begin{equation}
\uh(h_{C}) = \left\{ 
  \begin{array}{l l}
    1/(u-l) &  l \leq h_{C} \leq u \\
    0 & \text{otherwise}
  \end{array}\right.
\end{equation}

This simple approach to calibration already has two quite distinct
interpretations in a Bayesian setting when considered within the context of an
overall tree prior on all divergence times. One interpretation is that the
resulting marginal prior distribution on the calibrated divergence should obey
the tree process prior (e.g. Yule or birth-death) but be constrained to be
within the upper and lower bounds, so that the full calibrated tree prior,
$\fc(\cdot)$, is:

\begin{equation}
\fc(h,\RT|\Lambda)\propto \fBD(h,\RT|\Lambda) \uh(h_{C}),
\end{equation}

where $h$ represents the set of all divergence times, $\RT$ is the ranked tree
topology and $\Lambda$ represents the parameter(s) of the tree process
prior. The interpretation above was the only one available in the BEAST software
until recently \citep{heledDrummond2012}.  An alternative
``conditional-on-calibrated-node-ages'' interpretation is that the marginal
prior on the calibrated divergence should be uniform between the upper and lower
limits and the prior on the remaining divergence times should follow the tree
process prior, $\fBD(h,\RT|\Lambda)$, conditioned on the height of the calibrated
node \citep{yang2006bayesian}:

\begin{equation}
\fc(h,\RT|\Lambda) = \fBD(h \setminus h_{C},\RT|\Lambda) \uh(h_{C}),
\label{eq3}
\end{equation}

 There is a difference between these two prior
formulations regardless of whether the tree topology is known or estimated. In previous work
Heled and Drummond (2012) described how to efficiently compute the latter
formulation in the face of uncertainty in tree topology for arbitrary single-divergence calibration densities under the Yule
tree prior.

\subsection{\it Nested calibrations}

It has been routine in almost all treatments of phylogenetic calibration so far
to specify independent univariate priors for each calibrated divergence
time. However calibrated divergence times that are nested in the tree are
necessarily interdependent, such that the more recent calibrated divergence of a
nested pair must be younger than the older calibrated divergence. If the
specified calibration densities overlap than the resulting marginals of the
joint prior will necessarily differ from the specified calibration densities.
Nothing we present in this paper can correct for this. The correct solution to
this problem is to specify a joint prior on the calibrated nodes that obeys the
necessary condition that nested nodes are order statistics and therefore not
free to vary independently.

\subsection{\it The influence of calibrations on the tree topology prior}

\cite{heledDrummond2012} demonstrated that a natural interpretation of the
``conditional-on-calibrated-age'' construction of a calibrated tree prior
produces a distribution that is non-uniform on ranked topologies. However we
show in this paper that the tree prior can be decomposed into a prior on the
node ages (both calibrated and uncalibrated) and a prior on the set of possible
ranked histories. We show that this provides a means to compute a tree prior
rapidly if a uniform prior on ranked trees is chosen. We compare this approach
to a computational intensive alternative that weighs each ranked tree topology
by its probability conditional on the divergence times of the calibrated
nodes. The latter is a natural extension to our previous work to the case of
multiple calibrations and a birth-death process prior. However this extension turns out to be computational
expensive except for some special cases where a closed form formula exists. We therefore advocate the former
approach (that always applies a uniform prior to
ranked trees) as a practical alternative.

\section{Methods}

Consider the following notation:

\begin{description}
\item[$\nt$] Number of taxa.
\item[$\SRT$] The set of all ranked binary topologies on $\nt$ taxa. 
              We keep $\nt$ implicit to simplify the notation.
\item[$\RT$] A ranked tree ($\RT \in \SRT$).
\item[$h$] $= \{h_1, h_2, \cdots, h_{n-1}: h_i \ge h_{i+1} \ge 0 \}$, an ordered set
  of divergence ages.
\item[$g$] $=\langle h, \RT \rangle$, a time tree on $n$ taxa.
\item[$\G$] the space of all time trees.
\item[$\lambda$] the birth rate of the birth-death or Yule time tree prior.
\item[$\mu$] the death rate of the birth-death time tree prior.
\item[$\theta$] $=\langle \Omega, r \rangle$, a pair of parameter vectors, one
  for the substitution process $\Omega$ and one for the rates of the molecular
  clock, $r$.
\end{description}

\subsection{Posterior probability for Bayesian Inference}

Without calibration, the posterior probability of $(g,\Lambda,\theta)$ given a
sequence alignment, $D$ can be written:

\begin{equation}
f(g,\Lambda,\theta|D)= \frac{Pr\{D|g,\theta\}f(\theta)\fBD(g|\Lambda)f(\Lambda)}{Pr\{D\}}.
\end{equation}

The term $Pr\{D|g,\theta\}$ is the phylogenetic likelihood
\citep{felsenstein1981evolutionary}. The rates $r$ and divergence times $h$
combine to provide branch lengths in units of substitutions per site on the
edges of $\RT$. The term $\fBD(g|\Lambda)$ is the uncalibrated tree
prior and it can be readily factored in the following way:
\begin{align}\label{eqn:treepriorfact}
\begin{split}
\fBD(g | \Lambda)
= f(h | \Lambda) \Pr(\RT|\Lambda).
\end{split}
\end{align}

$f(h | \Lambda)$ is easy to compute for the pure birth (Yule) prior, birth-death
prior or any prior whose equivalence classes are defined entirely by the
divergence time order statistics. Under the Yule or birth-death prior {\it
  without calibrations}, $\Pr(\RT|\Lambda) = |\SRT|^{-1}$, is a uniform prior on
all ranked topologies. However this factorisation is no longer simple when
calibrations are introduced \citep{heledDrummond2012}, and so we must develop an
alternative approach to describing the calibrated tree prior in the following
sections, which we will call $\fc(\cdot)$ to distinguish from the uncalibrated
tree prior $\fBD(\cdot)$. Note that throughout the remaining sections the tree
priors are always conditional on $\Lambda$, but we suppress the conditioning in
the notation for the sake of clarity.

\subsection{Calibrated Birth-Death Density}

We introduce some extra notation for calibrations:

\begin{description}
\item[$\nc$] Number of calibration points.

\item[$\C$] Set of conditions on $\SRT$, typically clade monophyly
  constraints. $\C$ plays a part in the terms defined below, but since it is
  fixed in each case we mostly keep it implicit to make the equations easier to
  read.

\item[$\SRT_\C$] The subset of all ranked topologies for which $\C$ holds. 

\item[$\TR{}(\RT)$] $= (\TRI{1},\TRI{2},\cdots,\TRI{}{\nc})$, mapping a
  ranked tree to the ranks of the calibrated nodes. Typically those are
  the ranks of the monophyletic clades in $\C$, but $\TR{}$ may, for example,
  pick the rank of a clade's parent instead.

  We use two additional mappings which are a function of $\TR{}$. $\RA{}(\RT)
  = (\RAI{1},\RAI{2},\cdots,\RAI{}{\nc})$ is the mapping of calibration ranks
  into their sort order. For example, if $\TR{} = (3,1,4)$ then $\RA{} =
  (2,1,3)$ and if $\TR{} = (7,4,2)$ then $\RA{} = (3,2,1)$.  

  Also, $\AR{}(\RT) = (\ARI{1},\ARI{2},\cdots,\ARI{}{\nc}) =
  (\TRI{\RAI{1}},\TRI{\RAI{2}},\cdots,\TRI{\RAI{}{\nc}})$ are the ranks of
  the calibrated nodes sorted by age. For the two examples above we have
  respectively $\AR{} = (1,3,4)$ and $(2,4,7)$.

\item[$h_\RT$] $= (h_{\TRI{1}},h_{\TRI{1}},\cdots{},h_{\TRI{\nc}})$, the
  heights of the calibration points on a given ranked tree $\RT$. For
  convenience $g_\RT$ is the same as $h_\RT$ when $g = \langle h, \RT \rangle$. 

\item[$\uh(h_\RT)$] A $\nc$-dimensional calibration density. 

\end{description}

Figure \ref{fig:notation} illustrates the main elements of our notation on an
example tree with seven taxa and three calibrated sub-clades.

\newif\ifinlf
\inlftrue

\ifinlf
\begin{figure}[htp!]
\centering
\newcommand{\sss}{1}

\begin{tikzpicture}[scale=0.8,node distance = 1cm,auto,transform shape]

\tikzstyle{tip} = [text centered]
\tikzstyle{internal} = [thick, green]
\tikzstyle{calibrated} = [thick, red, fill]
\tikzstyle{line}=[draw,blue,thick]
\tikzstyle{hline}=[densely dotted, thick]
\tikzstyle{hl} = [text centered]

\node[tip] (a) {A};
\node[tip] (b) [right= .5 of a] {B} ;
\node[tip] (c) [right= .5 of b] {C};
\node[tip] (d) [right= .5 of c] {D};
\node[tip] (e) [right= .5 of d] {E};
\node[tip] (f) [right= .5 of e] {F};
\node[tip] (g) [right= .5 of f] {G};

\node (bc) [above right= 0.5 and .20 of b] {};
\node (abc) [above left= \sss and .75 of bc] {};
\node (abcd) [above right= 1.5 and \sss of abc] {};
\node (ef) [above right= \sss and 0.25 of e] {};
\node (efg) [above right= \sss and .5 of ef] {};
\node (abcdefg) [above right= 1 and 1.5 of abcd] {};

\draw[calibrated] (bc) circle (3pt);
\draw[internal] (abc) circle (3pt);
\draw[calibrated] (abcd) circle (3pt);
\draw[internal] (ef) circle (3pt);
\draw[calibrated] (efg) circle (3pt);
\draw[internal] (abcdefg) circle (3pt);


\draw[line] (bc) -- (b);
\draw[line] (bc) -- (c);
\draw[line] (abc) -- (a);
\draw[line] (abc) -- (bc);
\draw[line] (abcd) -- (d);
\draw[line] (abcd) -- (abc);
\draw[line] (ef) -- (e);
\draw[line] (ef) -- (f);
\draw[line] (efg) -- (ef);
\draw[line] (efg) -- (g);
\draw[line] (abcdefg) -- (abcd);
\draw[line] (abcdefg) -- (efg);

\node (h6) [left= 2 of bc] {};
\node (he6) [right= 7 of h6] {};

\draw[hline] (h6) -- (he6);

\node[hl] [left= .1 of h6] {\Large{$h_6$}};

\node (h5) [above= .25 of h6] {};
\node (he5) [right= 7 of h5] {};

\draw[hline] (h5) -- (he5);

\node[hl] [left= .1 of h5] {\Large{$h_5$}};

\node (h4) [above= .5 of h5] {};
\node (he4) [right= 7 of h4] {};

\draw[hline] (h4) -- (he4);

\node[hl] [left= .1 of h4] {\Large{$h_4$}};

\node (h3) [above= .25 of h4] {};
\node (he3) [right= 7 of h3] {};

\draw[hline] (h3) -- (he3);

\node[hl] [left= .1 of h3] {\Large{$h_3$}};

\node (h2) [above= 1 of h3] {};
\node (he2) [right= 7 of h2] {};

\draw[hline] (h2) -- (he2);

\node[hl] [left= .1 of h2] {\Large{$h_2$}};

\node (h1) [above= 1 of h2] {};
\node (he1) [right= 7 of h1] {};

\draw[hline] (h1) -- (he1);

\node[hl] [left= .1 of h1] {\Large{$h_1$}};

\end{tikzpicture}
\caption{For the tree above we have: $n=7$ taxa, the ranked topology
$\RT=$\txtr{(((a,(b,c):1):3,d):5,((e,f):2,g):4)} in NEWICK format, with internal
nodes marked by rank, $\C= (\{\txtr{e,f,g}\}, \{\txtr{b,c}\},\{\txtr{a,b,c,d}\})$,
$\nc=3$ calibrated nodes marked in red, and $\TR{}(\RT) = (3,6,2)$, $\RA{}(\RT)
= (2,3,1)$, $\AR{}(\RT) = (2,3,6)$ and $h_\RT = (h_3,h_6,h_2)$.}
\label{fig:notation}
\end{figure}
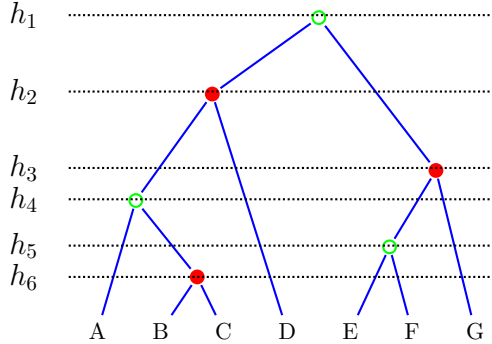
\fi

In BEAST, the calibrated tree prior has been defined as,
\begin{align}\label{eqn:beastmult}
\fc^{(\textsc{m})}(g) \bydef \fBD(g) \uh(h_\RT).
\end{align}
We shall call this the multiplicative prior, as designated by the superscript (M). While
multiplying the two densities creates some valid (unnormalized) prior density,
this tree prior fails to preserve the calibration density as the marginal prior
distribution of the calibrated nodes. That is, the marginal calibration density -- the
density obtained by integrating out the non-calibrated heights over all time
trees -- is {\bf not equal} to $\uh$.

In \cite{heledDrummond2012} we showed that it is easy in principle to preserve
the calibration marginal by scaling the prior with the conditional marginal
value, that is, the total density of all trees whose calibration times are
identical to the calibration times of $g$:

\begin{align}
\fM(x) = \int \limits_{\substack{g \in \G\\g_{\RT}=x}} \fBD(g) \,\D g.
\end{align}

The same general principle works for multiple calibrations:

\begin{align}\label{eqn:beastcond}
\fc(g) \bydef 
 \fBD(g) \frac{\uh(h_\RT)}{\fM(h_\RT)} .
\end{align}

The notation for describing the calibrated prior is challenging because
calibrated clade ages are not a simple subset of all ages. It may seem natural
to define the joint density by defining the tree prior density as the product of
conditional and calibration priors as done in Equation 3 of
\citep{yang2006bayesian} and mirrored in our own equation \ref{eq3}. But then
Yang and Rannala deal only with trees whose ranked topology is known. For
example, with this formulation one can easily forget that the space of possible
values for the uncalibrated nodes depends on the tree topology and the
calibrated nodes, and while this notational omission may be fine when the
topology is fixed, it should be explicit when dealing with the whole of tree
space. We think our notation is better suited for describing the properties of
the prior in the context of the full tree space.

The conditional prior in equation \eqref{eqn:beastcond} preserves the marginal
by construction. This is easy to see by writing down the marginal density
for $x$, a fixed vector of calibration values: 
{\small
\begin{align}\label{eqn:beastcondproof}
\int \limits_{\substack{g \in \G\\h_\RT = x}} \fBD(g) \frac{\uh(h_\RT)}
{\fM(h_\RT)} \,\D g = 
\int \limits_{\substack{g \in \G\\h_\RT = x}} \fBD(g) \frac{\uh(x)}
{\fM(x)} \,\D g  
=  \uh(x) \;\; \frac{\int \limits_{\substack{g \in \G\\h_\RT = x}} \fBD(g) \,\D g}
{\fM(x)} = 
\uh(x) .
\end{align}
}

However, the usefulness of this prior depends upon the computational cost of
evaluating $\fM(x)$ as part of the full posterior. In a few cases we can obtain
a simple formula and the cost is negligible, and for the rest we offer either
(a) a general algorithm for computing the marginal by iteration or (b) the
\textit{restricted conditional}, a fast alternate correction to be used when (a)
is too slow. The iterative approach is based upon the \textit{clade level
  partition}, which divides $\SRT_\C$ into disjoint subgroups whose marginal has
a closed form, and we shall discuss the details later.

The restricted conditional prior is defined as follows
\begin{align}\label{eqn:beastcond1}
\fc\rd(g = \langle \RT,h\rangle) \bydef 
\fBD(g) \frac{\uh(h_\RT)}{\fM\rd(h_\RT,\RT)} ,
\end{align}
Where 
\begin{align}
\fM\rd(x,\RT) = \int \limits_{h_{\RT}=x}
\fBD(\langle \RT,h \rangle) \,\D h.
\end{align}

Here the correction is defined as the marginal of the tree prior density when
keeping {\bf both} the topology and calibrated ages fixed. This is equivalent to
extending the approach taken by \cite{yang2006bayesian} to the case of an
unknown tree topology. Again, the marginal over tree space is preserved by
construction,

\begin{align}\label{eqn:beastcond1proof}
 \int \limits_{\substack{g = \langle \RT,h\rangle \in \G \\ h_\RT = x}}
\fBD(g) \frac{\uh(h_\RT)}{\fM\rd(h_\RT,\RT)} \,\D g  = 
 \sum_{\RT \in \SRT_\C} \;\;
\int \limits_{\substack{g = \langle \RT,h \rangle \\ h_\RT = x}}
\fBD(g) \frac{\uh(x)}{\fM\rd(x,\RT)} \,\D h  =  \nonumber \\
 \uh(x) \sum_{\RT \in \SRT_\C} \;\;
\frac{1}{\fM\rd(x,\RT)} \int \limits_{\substack{g = \langle \RT,h \rangle \\ h_\RT = x}}
{\fBD(g)} \,\D h = \uh(x) \sum_{\RT \in \SRT_\C} 1   \propto  \uh(x) .
\end{align}

\subsection{The Marginal Yule for Multiple Calibrations}

We start by showing how to decompose the Yule density of genealogy $g =
\langle\RT,\HE\rangle$ conditional on $\C$. 
The decomposition is based on separating the heights into $r+1$
\textit{levels}, where each level spans the range between two consecutive
calibration points.

The Yule density 
\begin{equation}
\fBD(h|\lambda) = \frac{1}{| \SRT_\C |}
          n! e^{-\lambda h_1} \prod_{i=1}^{n-1} \lambda e^{-\lambda h_i}
\end{equation}
is factored using the two propositions below.

{
\small
\begin{description}
\item[Proposition I:] 
\begin{equation}
\begin{aligned}
\int_a^b {\scriptstyle \D x_1} \int_a^{x_1} {\scriptstyle \D x_2} 
\int_a^{x_2}  {\scriptstyle \D x_3} {\scriptstyle \cdots}
\int_a^{x_{k-1}} {\scriptstyle \D x_k} \;\; 
\lambda e^{-\lambda x_1} \lambda e^{-\lambda x_2} 
\cdots \lambda e^{-\lambda x_k} = \\
\frac{1}{k!} \left[ \int_a^b \lambda e^{-\lambda x_1} \,\D x_1 \right] 
\left[ \int_a^b \lambda e^{-\lambda x_2} \,\D x_2 \right] {\scriptstyle \cdots}
\left[ \int_a^b \lambda e^{-\lambda x_k} \,\D x_k \right] = \\
  \frac{1}{k!} \left( e^{-\lambda b} - e^{-\lambda a} \right)^k 
\end{aligned}
\end{equation}

\item[Proposition II:] 
\begin{equation}
\begin{split}
& \int_a^\infty {\scriptstyle \D x_0} \int_a^{x_0} {\scriptstyle \D x_1} 
\int_a^{x_1} {\scriptstyle \D x_2} {\scriptstyle \cdots}
\int_a^{x_{k-1}} {\scriptstyle \D x_k} \;\; 
  {\scriptstyle \lambda e^{-2 \lambda x_0} \lambda e^{-\lambda x_1} \cdots \lambda e^{-\lambda
    x_k} } \\
& \quad \text{by proposition I} \\
& = \int_a^\infty \lambda e^{-2 \lambda x_0} \frac{1}{k!} \left( e^{-\lambda a} -
  e^{-\lambda x_0} \right)^k   \,\D x_0 \\
& \quad \text{by equation \eqref{eqn:intkabove}} \\
& = \frac{1}{(k+2)!} e^{-(k+2)\lambda a}.
\end{split}
\end{equation}
\end{description}
}

Proposition I gives the contribution of internal nodes located between two
consecutive calibration points with ages $a$ and $b$. Proposition II gives the
contribution of nodes older than the last calibration point.

When the calibration values are fixed to $x = ( x_1,x_2,\cdots,x_\nc )$, the
contribution of the ranked topology $\RT$ is

\begin{equation}
\begin{split}
\fM(x,\RT) = & 
  \int_{\bar{x}_1}^\infty \D \HE_1
  \int_{\bar{x}_1}^{\HE_1} \D \HE_2
     \cdots
  \int_{\bar{x}_1}^{\HE_{\ARI{1}-3}} \D \HE_{\ARI{1}-2} 
  \int_{\bar{x}_1}^{\HE_{\ARI{1}-2}} \D \HE_{\ARI{1}-1} \\
& \int_{\bar{x}_2}^{\bar{x}_1} \D \HE_{\ARI{1}+1} 
  \int_{\bar{x}_2}^{\HE_{\ARI{1}+1}} \D \HE_{\ARI{1}+2} 
     \cdots 
   \int_{\bar{x}_2}^{\HE_{\ARI{2}-3}} \D \HE_{\ARI{2}-2} 
   \int_{\bar{x}_2}^{\HE_{\ARI{2}-2}} \D \HE_{\ARI{2}-1} \\
& \int_{\bar{x}_3}^{\bar{x}_2} \D \HE_{\ARI{2}+1}
  \int_{\bar{x}_3}^{\HE_{\ARI{2}+1}} \D \HE_{\ARI{2}+2}
     \cdots 
  \int_{\bar{x}_3}^{\HE_{\ARI{3}-3}} \D \HE_{\ARI{3}-2}
  \int_{\bar{x}_3}^{\HE_{\ARI{3}-2}} \D \HE_{\ARI{3}-1}\\
& \quad \quad \quad  \dotsi \\
& \int_{0}^{\bar{x}_{\nc}} \D \HE_{\ARI{\nc}+1}
   \int_{0}^{\HE_{\ARI{\nc}+1}} \D \HE_{\ARI{\nc}+2}
    \cdots
   \int_{0}^{\HE_3} \D \HE_2
   \int_{0}^{\HE_2} \D \HE_1
         f_Y(\langle\RT,\HE\rangle) .
\end{split}
\end{equation}

The above uses $\bar{x} = (x_{\RAI{1}},x_{\RAI{2}},\cdots,x_{\RAI{\nc}})$, the
calibration height sorted by age. Now, let $c_i$ be the number of internal nodes
in each level, $c_i = \ARI{{i+1}} - \ARI{i} - 1 \;\; (0 \le i \le \nc)$, and for
convenience let $\ARI{0} = 0$ and $\ARI{\nc + 1} = \nt$. Using propositions I
and II we get

\begin{align}\label{eqn:topmarginal}
\begin{split}
  \fM(x,\RT) = &
 \frac{\nt{}!}{| \SRT_\C |} 
\frac{e^{- (c_0 + 1) \lambda \bar{x}_1}}{(c_0+1)!}  \\
& \frac{\lambda e^{- \lambda \bar{x}_1}}{c_1!} \left( e^{-\lambda \bar{x}_2} - e^{-\lambda
    \bar{x}_1} \right)^{c_1} 
 \frac{\lambda e^{- \lambda \bar{x}_2}}{c_2!} \left( e^{-\lambda \bar{x}_3} - e^{-\lambda
    \bar{x}_2} \right)^{c_2} 
  \cdots 
  \frac{\lambda e^{- \lambda \bar{x}_{\nc}}}{c_{\nc}!} \left( e^{-0} - e^{-\lambda \bar{x}_{\nc}}
\right)^{c_{\nc}} \\
= & \left[ \frac{\nt{}!} {|\SRT_\C|} \prod_{k=1}^{\nc} \lambda e^{-\lambda
    \bar{x}_k} \right] 
\frac{e^{- (c_0 + 1) \lambda \bar{x}_1}}{(c_0+1)!} 
 \;\;\;     \prod_{k=1}^{\nc} \frac{\left(  e^{-\lambda \bar{x}_{k+1}} - e^{-\lambda \bar{x}_{k}}
  \right)^{c_k}}{c_k!} .
\end{split}
\end{align}

The marginal density is the sum over all \textbf{valid} topologies
\begin{equation}\label{eqn:marginal}
{\fM}(x) = \sum_{\substack{\RT \in \SRT_\C \\ \RA(\RT) = \RA(x)}} \fM(x,\RT) .
\end{equation}

To be valid, the order of calibration points by age has to be compatible with
$x$.

While explicitly summing over all topologies is not feasible, evaluating the sum
is possible by partitioning $\SRT_\C$ into $\{\SRT_\C^1,\SRT_\C^2,\cdots\}$,
where $\RT_1,\RT_2 \in \SRT_\C^k \implies \TR(\RT_1) = \TR(\RT_2)$. That is,
topologies in the same partition have the same number of internal nodes in each
\textit{level}. Since equation \eqref{eqn:topmarginal}
depends only on those counts ($c_i$) and not on the exact ranking, we have
$f_T(x,\RT_1) = f_T(x,\RT_2)$ for two topologies in the same partition. Finally,

\begin{equation}\label{eqn:marginal-split}
{\fM}(x)  = \sum_{\substack{\RT \in \SRT_\C \\ \RA(\RT) = \RA(x)}} \fM(x,\RT) = 
\sum_{\substack{k \\\RA(\RT_k) = \RA(x)} } \vert \SRT_\C^k \vert \fM(x,\RT_k) 
\end{equation}

where $\RT_k$ is any topology in $\SRT_\C^k$.

\subsection{The Marginal Birth-Death Prior for Multiple Calibrations}

\newif\ifwithrho
\withrhotrue

The birth-death process starts with a single species, and evolves over time
through existing species giving birth (splitting) to new species at constant
rate $\lambda$ and dying (unobserved) at constant rate $\mu$
\citep{kendall1948generalized}. While this characterisation is unique, there are
several versions of the prior which differ in their start and end conditions.
BEAST uses the birth-death-sampling$_\rho$ process, which assumes a uniform
distribution $[0,\infty]$ on the time of the tree origin, and that the leafs of
the tree are sampled with probability $\rho$ to obtain exactly $\nt$ taxa. The
density for this prior is given in 
\ifwithrho 
equation (5) of \citep{stadler2009incomplete}: 
\else 
section 6.1 of \citep{gernhard2008conditioned}: 
\fi

\ifwithrho
{\scriptsize
\begin{equation}
\fBD(h \vert \lambda,\mu,\rho) = 
n! (\rho\lambda)^{n-1} \frac{(\lambda - \mu) e^{-(\lambda - \mu)h_1} }
{\rho\lambda + (\lambda(1-\rho) -\mu) e^{-(\lambda - \mu)h_1} } \prod \limits_{i=1} ^{n-1} 
\frac{(\lambda - \mu)^2 e^{-(\lambda - \mu)h_i} }
{\left( \rho\lambda + (\lambda(1-\rho) -\mu) e^{-(\lambda - \mu)h_i} \right) ^ 2}.
\end{equation}
}
\else
\begin{equation}
\fBD(h \vert \lambda,\mu) = 
n! \lambda{}^{n-1}(\lambda - \mu) \frac{e^{-(\lambda - \mu)h_1} }
{\lambda - \mu e^{-(\lambda - \mu)h_1} } \prod \limits_{i=1} ^{n-1} 
\frac{(\lambda - \mu)^2 e^{-(\lambda - \mu)h_i} }
{\left( \lambda - \mu e^{-(\lambda - \mu)h_i} \right) ^ 2}.
\end{equation}
\fi

We obtain the marginal for the birth-death process using exactly the same
procedure as described for the Yule, but using the birth-death analogous for
propositions I and II. We use the following definitions for convenience:

\ifwithrho
\begin{align}
\lambda' & =  \rho\lambda \\
\mu' & =  \mu - \lambda(1-\rho) \\
q(t) & =  \frac{\lambda - \mu}{ \lambda' - \mu' e^{-(\lambda - \mu)t} } \\
q_1(t) & =  e^{-(\lambda - \mu) t} q(t) . \\
p_1(t) & =  q_1(t) q(t)
\end{align}
\else
\begin{align}
p_1(t) = & \frac{(\lambda - \mu)^2 e^{-(\lambda - \mu) t} }
{\left( \lambda - \mu e^{-(\lambda - \mu)t} \right) ^ 2}. \\
q(t) = & \frac{\lambda - \mu}{\lambda - \mu e^{-(\lambda - \mu)t} } \\
q_1(t) = & e^{-(\lambda - \mu) t} q(t) .
\end{align}
\fi

$p_1(t)$ is the probability that a lineage leaves one descendant after time $t$,
which is easy to integrate
\ifwithrho
\begin{equation}
P_1(t) = \int p_1(t) dt = -\frac{q(t)}{\mu'} ,
\end{equation}
\else
\begin{equation}
P_1(t) = \int p_1(t) dt = -\frac{q(t)}{\mu} ,
\end{equation}
\fi

and gives us the birth-death equivalent of proposition I:
\begin{equation}
\begin{aligned}
\int_{a}^b \D x_1 \int_{a}^{x_1} \D x_2  {\scriptstyle \cdots}
\int_{a}^{x_{k-1}} \D x_k  \prod_{i=1}^k p_1(x_k)  = 
\frac{1}{k!} \left ( P_1(b) - P_1(a) \right )^{k} .
\end{aligned}
\end{equation}

For proposition II we have
\ifwithrho
\begin{equation}
\begin{aligned}
\int_{a}^\infty \D x_0 \int_{a}^{x_0} \D x_1  {\scriptstyle \cdots}
\int_{a}^{x_{k-1}} \D x_k \,\,
   q_1(x_0) \prod_{i=0}^k p_1(x_k)  = 
\frac{\lambda'^{-(k+1)}}{(k+2)!} q_1(a)^{k+2}
\end{aligned}
\end{equation}
\else
\begin{equation}
\begin{aligned}
\int_{a}^\infty \D x_0 \int_{a}^{x_0} \D x_1  {\scriptstyle \cdots}
\int_{a}^{x_{k-1}} \D x_k \,\,
   q_1(x_0) \prod_{i=0}^k p_1(x_k)  = \\
\frac{\lambda^{-(k+1)}}{(k+2)!} q_1(a)^{k+2}
\end{aligned}
\end{equation}
\fi

Which is proved in the appendix. For the critical case $\lambda = \mu$ we take
the limit and use
\ifwithrho
$q_1(t) = \frac{1}{1+\lambda' t}$
\else
$q_1(t) = \frac{1}{1+\lambda t}$
\fi
 in the formulas above.

\subsection{Partitioning and Counting}

To evaluate the marginal (equation \eqref{eqn:marginal-split}) we need to
establish a valid partitioning and count the number of ranked topologies in each
partition. Ideally, the partition would be the smallest possible, that is
$\RT_1,\RT_2 \in \SRT_\C^k \iff \TR(\RT_1) = \TR(\RT_2)$. Unfortunately, we were
unable to derive a counting formula under this constraint and instead use the
clade level partition, a refinement based on the number of lineages per level
inside each calibrated clade. Formally, let $r(\RT) = \{r_1,r_2,\cdots,r_K\}$
where $r_j= (r_{j0},r_{j1},\cdots,r_{jK})$ and $r_{jk}$ is the number of
sub-clades of the $k^{th}$ calibration point whose rank is smaller than
$\TRI{j}$. Since $1 + \sum_k r_{jk} = \TRI{j}$, the equivalence classes induced
by $r$ are a refinement of the ones induced by $\TR(\cdot)$. Furthermore, we can
count the number of topologies in each class by using two generic combinatorial
principles: first, the number of ways for lineages to coalesce in each level is
independent of other levels, so the product of counts of all levels gives the
total number of topologies. Second, when $n = n_1 + n_2 + \cdots + n_j$ lineages
enter a level and are reduced to $k = 1 + k_2 + \cdots + k_j$, where lineages
can coalesce only within their group ($n_i \rightarrow k_i$) and the root of the
first group is calibrated ($k_1=1$), the total number of ranked ways is
$\binom{n-k-1}{n_1-2,n_2-k_2,\cdots,n_j-k_j} \prod_{i=1}^j \NR_{n_i}^{k_i}$.

$\NR_n^k$ is the number of ranked ways $n$ lineages can coalesce to $k$
(equation \eqref{eq:rnk}), and for convenience $\NR_n = \NR_n^1$.

Figure \ref{fig:counting} illustrates the counting procedure on a small example
tree.

\ifinlf
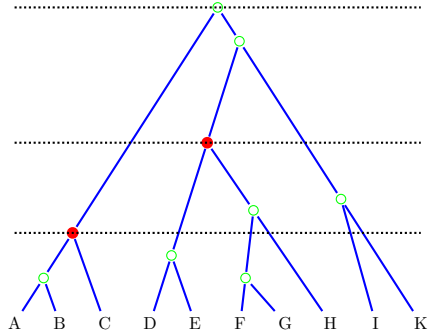
\begin{figure}[htp!]
\centering
\begin{tikzpicture}[scale=0.6,node distance = .5cm,auto,transform shape]


\tikzstyle{tip} = [text centered]
\tikzstyle{internal} = [green]
\tikzstyle{calibrated} = [thick, red, fill]
\tikzstyle{line}=[draw,blue,thick]
\tikzstyle{hline}=[densely dotted, thick]

\node[tip] (c1) at (0,0) {A};
\node[tip] (c2) at (1,0) {B};
\node[tip] (c4) at (2,0) {C};
\node[tip] (c6) at (3,0) {D};
\node[tip] (c7) at (4,0) {E};
\node[tip] (c9) at (5,0) {F};
\node[tip] (c10) at (6,0) {G};
\node[tip] (c12) at (7,0) {H};
\node[tip] (c15) at (8,0) {I};
\node[tip] (c16) at (9,0) {K};
\node (c0) at (4.5,7.0) {};
\node (c3) at (0.642857142857,1.0) {};
\node (c5) at (1.28571428571,2.0) {};
\node (c8) at (3.47571428571,1.5) {};
\node (c11) at (5.11714285714,1.0) {};
\node (c13) at (5.29285714286,2.5) {};
\node (c14) at (4.26857142857,4.0) {};
\node (c17) at (7.23214285714,2.75) {};
\node (c18) at (4.98214285714,6.25) {};
\draw[internal] (c0) circle (3pt);
\draw[internal] (c3) circle (3pt);
\draw[calibrated] (c5) circle (3pt);
\draw[internal] (c8) circle (3pt);
\draw[internal] (c11) circle (3pt);
\draw[internal] (c13) circle (3pt);
\draw[calibrated] (c14) circle (3pt);
\draw[internal] (c17) circle (3pt);
\draw[internal] (c18) circle (3pt);
\draw[line] (c1) -- (c3);
\draw[line] (c2) -- (c3);
\draw[line] (c3) -- (c5);
\draw[line] (c4) -- (c5);
\draw[line] (c5) -- (c0);
\draw[line] (c6) -- (c8);
\draw[line] (c7) -- (c8);
\draw[line] (c8) -- (c14);
\draw[line] (c9) -- (c11);
\draw[line] (c10) -- (c11);
\draw[line] (c11) -- (c13);
\draw[line] (c12) -- (c13);
\draw[line] (c13) -- (c14);
\draw[line] (c14) -- (c18);
\draw[line] (c15) -- (c17);
\draw[line] (c16) -- (c17);
\draw[line] (c17) -- (c18);
\draw[line] (c18) -- (c0);
\draw[hline] (0,7.0) -- (9,7.0);
\draw[hline] (0,2.0) -- (9,2.0);
\draw[hline] (0,4.0) -- (9,4.0);

\end{tikzpicture}
\caption{To count the number of ranked topologies for the tree above, we
  multiply the counts in the 3 levels. In the lowest level we have 3 lineages
  reducing to 1 (root of lowest calibration), 5 lineages reducing to 3 and 2
  free lineages not reducing. So, the total number of topologies is $\NR_3^1
  \NR_5^2 \NR_2^2 \binom{2+5+2 - (1+3+2)}{1,2,0} = 3 \times (10\times 6) \times
  1 \times \frac{3!}{1!2!0!} = 540$ Note that in the binomial we use one less
  lineage (2 instead of 3) for the calibrated clade, since its position as root
  is fixed.  In the second level we have 3 lineages reducing to 1, and 3 free
  lineages reducing to 2, giving $\NR_3^1 \NR_3^2 \binom{2+3 - (1+2)}{1,1} = 3
  \times 3 \times \frac{2!}{1!1!} = 9$ and in the last level 3 lineages to 1 in
  3 ways. So, the total number is $540 \times 9 \times 3 = 12960$.  }
\label{fig:counting}
\end{figure}
\fi

The use of the clade level partition has an interesting consequence, which
relates to the second property of the conditional prior, namely that trees are
``Yule-like'' (or ``birth-death-like'') conditional on the calibrated ages. This
means that the density ratio of trees with equal calibrated ages is the same as
their density ratio under the uncalibrated tree prior alone (equation (3) in
\cite{heledDrummond2012}). This condition is relaxed for the restricted
conditional prior, where by construction this ratio equality is true only for
trees with the same ranked topology. However, since the marginal (equation
\eqref{eqn:topmarginal}) depends only on the number of lineages between levels
and not on the exact ranked topology, the space in which each tree is
``birth-death-like'' is in fact larger, containing all trees in the same
partition.
 
\subsection{Enumerating Ranked Topologies Classes}

Here we explain the procedure for explicitly enumerating all the elements of the
clade level partition. The enumeration is based upon combining several
iterators, one for every calibrated clade, which return the number of lineages
in each level of that clade. Those counts are used to compute the marginal as
explained in the previous section. The calibrated nodes and the root of the
tree, which define the levels, are not included in the counts. We show the
working via an example; the interested reader should consult the source code for
the very low level details. The iterator is built from the product of $\nc+1$
per-clade iterators, one for each calibration and one for the ``free'' lineages
outside any calibrated clade. In fact, each calibrated clade is potentially
composed of several calibrated sub-clades and some free lineages, and the
iterator for the clade handles the free lineages and the surviving lineage from
the root of each calibrated sub-clade. Figure \ref{fig:figxx} gives an example
with three calibrated clades, $N$ with $n+2$ lineages, nested inside $M$ with
$n+m+3$ lineages, and $L$ with $l+2$ lineages. The uppercase letters are the
clade name, and the lowercase letter gives the number of additional lineages.

\ifinlf
\begin{figure}[htp!]
\centering
\includegraphics[width=6cm]{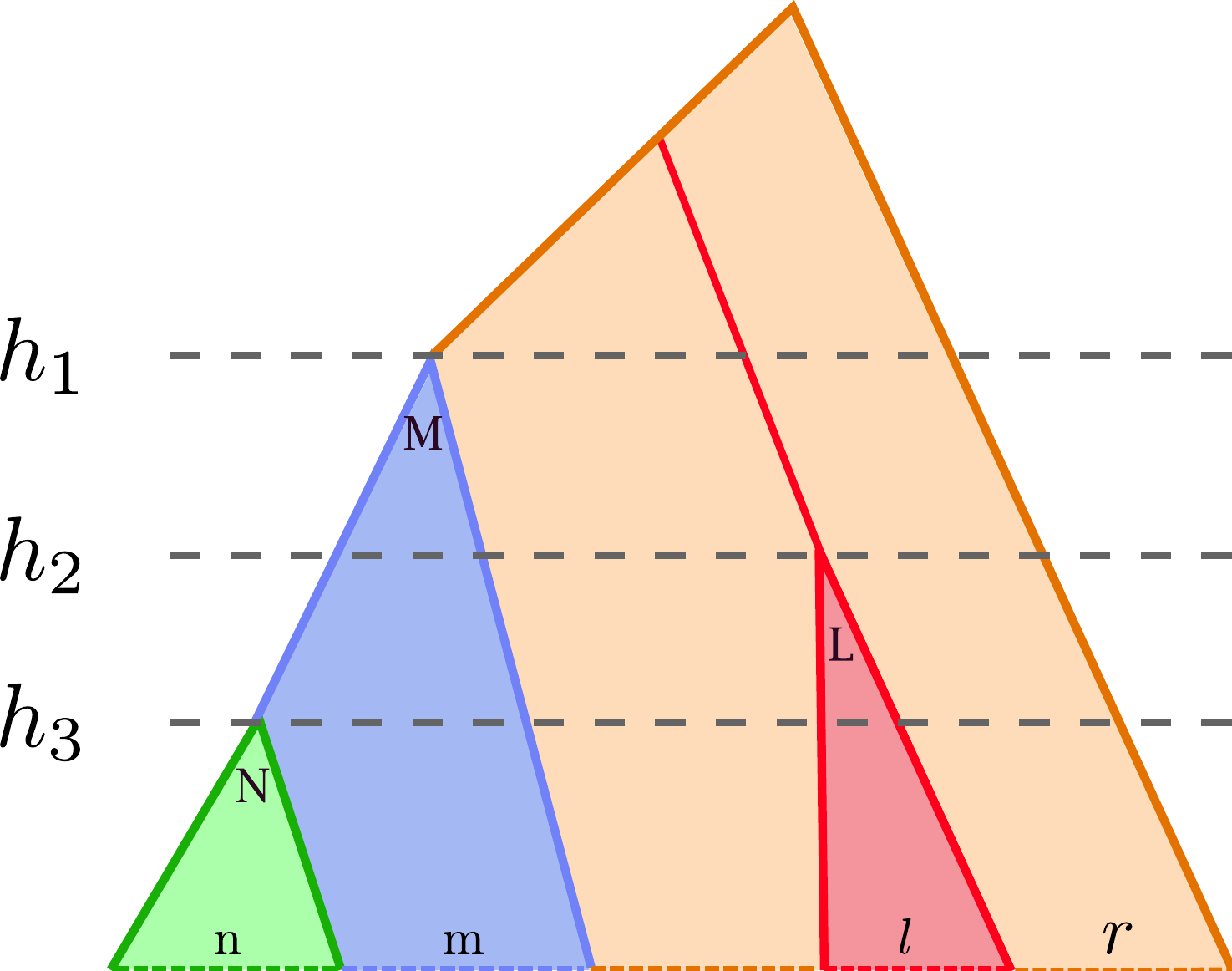}
\caption{The calibrated clades are N (green), M (blue) and L (red), with respectively
$n+2$, $n+2+m+1$ and $l+2$ taxa. In addition we have $r$ ``free'' taxa in
orange, which can coalesce between themselves and with the roots of M and
L. There are 4 levels associated with the 3 calibration nodes, separated by the
root ages of the calibrated clades.}
\label{fig:figxx}
\end{figure}
\fi

In addition there are $r \ge 0$ free lineages, for a total of $n+m+l+5+r$
lineages in the tree. The $m+1$ lineages of $M$ not in $N$ coalesce on the way
to the clade root with each other and with the roots of the nested clades, in
this case $N$. The $r$ free lineages coalesce with the roots of L and M on the
way up, and the internal nodes can be in any of the four levels.

Since there are 3 calibrations there are 4 levels, separated by the dashed
lines, and each per-clade iterator returns 4 numbers. The iterator of $N$ is
trivial, always returning $(n,0,0,0)$, since its root defines the first
level. The iterator for $L$ is simple too, since the lineages can coalesce only
in the first two levels and there are no free lineages. The iterator returns
$(l,0,0,0),(l-1,1,0,0),\dots (0,l,0,0)$.

The iterator for $M$ takes care of $m+1$ free lineages which can coalesce in the
first 3 levels. The iterator returns
$(m,0,0,0),(m-1,1,0),(m-1,0,1),(m-2,2,0),(m-2,1,1),(m-2,0,2)\dots,(0,0,m)$.
Basically, the iterator first returns all the cases with $m$ internal nodes in
the first level, then all cases with $m-1$ internal nodes in the first level,
and so on. The same pattern holds (recursively) for the rest of the levels.

The last iterator takes care of the $r$ free lineages and the surviving lineages
of any sub-clade, here the roots of M and L. In this example this iterator is
only necessary if $r>0$, as otherwise there are only two lineages left to deal
with. While the internal nodes can be in any of the 4 levels, there are some
restrictions. In general these restrictions can be quite involved. In this
example, the restrictions arise because the enclosing clade (here the root of
the tree) has more than one monophyletic sub-clade. As a result we always have
at least three lineages above $h_2$, and since only 2 lineages coalesce at the
root, the excess has to coalesce in the top 2 levels.  So, the iterator returns
$(r-1,0,1,0),(r-1,0,0,1),(r-2,1,1,0)\dots,(0,0,0,r)$, filling up lower levels
first as before, while keeping at least one event in the top two.

\section{Results}

\subsection{Calibrating the Parent of One Monophyletic Clade}
\label{poc}

Sometimes the calibration information is about the time a particular clade (say
a genera, or a species which is divided into sub-species) separated from other
lineages in the tree. For a single lineage, the density is given in
\cite{heledDrummond2012}

\begin{equation}
\fM(x) = 2 \lambda e^{-2 \lambda x}.
\end{equation}

Note that the parent age is equal to the (pendant) branch length, and in fact
$\fM(x)$ is the distribution of the branch length when conditioning on the
number of leaves. Furthermore, since this holds for any branch, we can derive a
mean of $^1/_{2 \lambda}$, which reproduces a result discussed by
\cite{steel2010expected}.

The result can be generalized to any clade $C$ of size $n$. In that case $\SRT_\C$
to be the set of all genealogies of $n+l$ taxa with a monophyletic clade on $n$
taxa ($n>1$ and $l>0$) (Figure \ref{fig:parentOfClade}).

\ifinlf
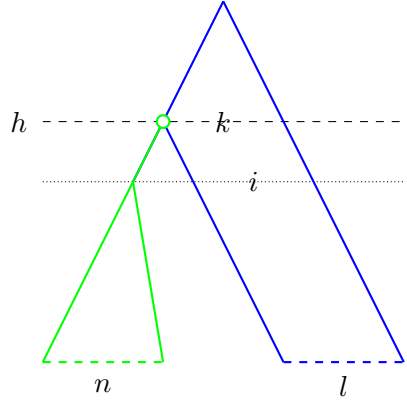
\begin{figure}[htp!]
\centering
\begin{tikzpicture}[scale=0.8]

\node at (-0.4, 4) {$h$};
\draw[dashed] (0,4) -- (6, 4);

\draw[densely dotted] (0,3) -- (6, 3);

\node at (1.0, -0.4) {$n$};
\node at (5.0, -0.4) {$l$};

\node at (3.5, 3) {$i$};
\node at (3, 4) {$k$};


\draw[thick, blue] (1.5,3) -- (3,6);
\draw[thick, blue] (4,0) -- (2,4);
\draw[thick, blue] (6,0) -- (3,6);
\draw[thick, blue, dashed] (4,0) -- (6,0);


\draw[thick, green] (0,0) -- (2,4);
\draw[thick, green, dashed] (0,0) -- (2,0);
\draw[thick, green] (2,0) -- (1.5,3);

\fill[white] (2,4) circle (3pt);
\draw[thick, green] (2,4) circle (3pt);

\end{tikzpicture}
\caption{Parent of monophyletic clade of size $n$.}
\label{fig:parentOfClade}
\end{figure}
\fi

We partition $\SRT_\C$ so that $\SRT_\C^k$ contains all genealogies containing
$k+1$ surviving lineages at $h$, the age of the calibrated parent. By the second
counting principle, there are $\NR_n \NR_l^i \binom{n-2+l-i}{n-2}$ ranked ways
for lineages to coalesce in the first level, $\NR_i^{k+1}$ ways for $i$ lineages
to reduce to $k+1$ in the second level, and then one of the $k+1$ coalesce with
the parent of $C$. Then $k+1$ lineages coalesce to the root, giving

\begin{equation}
\begin{split}
|\SRT_\C^k| & =  
\sum_{i=k+1}^l \NR_n \NR_l^i \binom{n-2+l-i}{n-2} (k+1) \NR_i^{k+1} \NR_{k+1} \\
& = (k+1) \binom{n-2 + l-k}{n-1} \NR_n \NR_l .
\end{split}
\end{equation}

The total number of ranked trees in $\SRT_\C$ is
\begin{equation}
|\SRT_\C| = \sum_{k=0}^{l-1} |\SRT_\C^k| = \binom{l+n}{l-1} \NR_l \NR_n .
\end{equation}

Putting it all together,

\begin{equation}
\begin{split}
\fM(x)  = & \frac{1}{|\SRT_\C|} \sum_{k=0}^{l-1} 
|\SRT_\C^k| (n+l)! \\
& \quad \quad e^{-\lambda x} \frac{e^{-(k+1)\lambda x}}{(k+1)!} 
\frac{(1 - e^{-\lambda x})^{n+l-k-2}}{(n+l-k-2)!} \\
 = & n(n+1) \sum_{k=0}^{l-1} \binom{l-1}{k} 
     e^{-(k+2)\lambda x}  (1 - e^{-\lambda x})^{n+l-(k+2)} \\
 = & n(n+1) \lambda e^{-2 \lambda x} \left(1-e^{-\lambda x}\right)^{n-1} .
\end{split}
\end{equation}

Note that the marginal does not depend the size of the tree, just on the size of
the calibrated clade.

\subsection{Calibrating Two Nested Clades}

Here we give the marginal density for two nested clades. When the enclosing
clade is the root (Figure \ref{fig:rootAndClade}), the marginal is

{\scriptsize
\begin{equation}
\label{eqn:rootandclade}
\begin{split}
\fM(h_0,h | n,m) = & (n - 1)n(n + 1) 
\lambda^2 e^{-\lambda (h + 2 h_0)} 
  ( 1-e^{-\lambda h})^{n-2} (1 - e^{-\lambda h_0})^{m-3} \\
&  \left [ 1 + 2(m-1) e^{-\lambda h} - 2 m e^{-\lambda h_0} - \right . 
     m(m-1) e^{-\lambda( h_0 + h)} + 
  \left . \binom{m-2}{2} e^{-2 \lambda h} +  \binom{m}{2} e^{-2 \lambda x0} \right ] ,
\end{split}
\end{equation}
}

\ifinlf
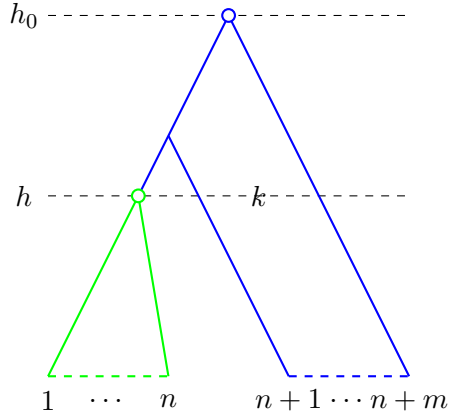
\begin{figure}[htp!]
\centering
\begin{tikzpicture}[scale=0.8]

\node at (-0.4,6) {$h_0$};
\node at (-0.4, 3) {$h$};
\node at (3.5, 3) {$k$};

\draw[dashed] (0,6) -- (6, 6);
\draw[dashed] (0,3) -- (6, 3);

\node at (0.0, -0.4) {1};
\node at (1.0, -0.4) {$\dots$};
\node at (2.0, -0.4) {$n$};
\node at (4.0, -0.4) {$n+1$};
\node at (5.0, -0.4) {$\dots$};
\node at (6.0, -0.4) {$n+m$};


\draw[thick, blue] (1.5,3) -- (3,6);
\draw[thick, blue] (4,0) -- (2,4);
\draw[thick, blue] (6,0) -- (3,6);
\draw[thick, blue, dashed] (4,0) -- (6,0);


\draw[thick, green] (0,0) -- (1.5,3);
\draw[thick, green, dashed] (0,0) -- (2,0);
\draw[thick, green] (2,0) -- (1.5,3);

\fill[white] (1.5,3) circle (3pt);
\draw[thick, green] (1.5,3) circle (3pt);

\fill[white] (3,6) circle (3pt);
\draw[thick, blue] (3,6) circle (3pt);

\end{tikzpicture}
\caption{  Monophyletic clade of size $n$ and root with $n+m$ taxa.}
\label{fig:rootAndClade}
\end{figure}
\fi

And when the enclosing clade is proper (Figure \ref{fig:twonested}), it is
{\scriptsize
\begin{equation}
\label{eqn:twonested}
\begin{split}
\fM(h_1,h_2 | n,m) = & ^1/_2 (n - 1)n(n + 1)(n+1+m)
\lambda^2  
 e^{-\lambda (h_2 + 3 h_1)} (1 - e^{-\lambda h_2})^{n-2} ( 1-e^{-\lambda h_1})^{m-3}  \\
& \left [ 1 - 2 m e^{-\lambda h_1} + 2(m-1) e^{-\lambda h_2} - \right . 
  m (m-1) e^{-\lambda(h_2 + h_1)} + 
 \left . \binom{m+1}{2} e^{-2\lambda h_1} + \binom{m-1}{2} e^{-2\lambda h_2}
\right ] .
\end{split}
\end{equation}
}

\ifinlf
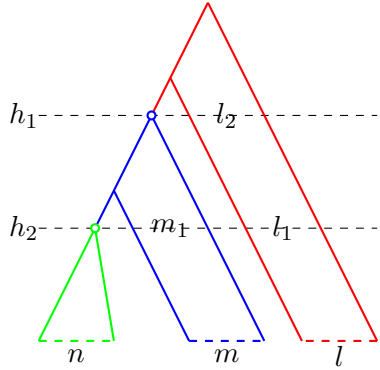
\begin{figure}[htp!]
\centering
\begin{tikzpicture}[scale=0.5]

\node at (-0.4,6) {$h_1$};
\node at (-0.4, 3) {$h_2$};
\draw[dashed] (0,6) -- (9, 6);
\draw[dashed] (0,3) -- (9, 3);

\node at (3.5, 3) {$m_1$};
\node at (6.5, 3) {$l_1$};
\node at (5, 6) {$l_2$};

\node at (1.0, -0.4) {$n$};
\node at (5.0, -0.4) {$m$};
\node at (8.0, -0.4) {$l$};


\draw[thick, blue] (1.5,3) -- (3,6);
\draw[thick, blue] (4,0) -- (2,4);
\draw[thick, blue] (6,0) -- (3,6);
\draw[thick, blue, dashed] (4,0) -- (6,0);


\draw[thick, green] (0,0) -- (1.5,3);
\draw[thick, green, dashed] (0,0) -- (2,0);
\draw[thick, green] (2,0) -- (1.5,3);

\draw[thick, red] (3,6) -- (4.5,9);
\draw[thick, red] (4.5,9) -- (9,0);
\draw[thick, red, dashed] (9,0) -- (7,0);
\draw[thick, red] (3.5,7) -- (7,0);

\fill[white] (1.5,3) circle (3pt);
\draw[thick, green] (1.5,3) circle (3pt);

\fill[white] (3,6) circle (3pt);
\draw[thick, blue] (3,6) circle (3pt);

\end{tikzpicture}
\caption{Two nested monophyletic clades of size $n$ and $n+m$ taxa in a $n + m
  + l$ taxa tree ($l > 0$).}
\label{fig:twonested} 
\end{figure}
\fi

See the appendix for additional details on the derivation of those formulas.

\subsection{Placing Additional Monophyly Constraints}

It is important to keep in mind that placing additional constraints can
invalidate the closed form equations for the marginal. However, it still may be
possible to obtain formula for the full set of constraints. For example, the
marginal density for a monophyletic clade of size $n$ in a $n+m+1$ taxa tree
with an outgroup can be obtained by integrating out $h_2$ in equation
\eqref{eqn:twonested} and is equal to

{ \small{
\begin{equation}
\begin{split}
\fM(x) =  & \frac{(n - 1)n(n+1)(n+m+1)}{m(m+1)(m+2)} 
 \lambda e^{-\lambda x} (1-e^{-\lambda x})^{n - 2} \\
& \left(1 - (1 - e^{-\lambda x})^{m+2}
   - (m+2) e^{-\lambda x} + \binom{m+2}{2} e^{-2\lambda x} \right) ,
\end{split}
\end{equation}
}
}
which is not equal to the marginal for the same sized tree where the
monophyly on the $n+m$ clade is not enforced.

We can also derive the marginal in some cases which are not covered by the
standard construction (root ages of monophyletic clades and no extra
constraints). For example, take the *BEAST analysis performed as part of the
investigation of determining the Pipid root \citep{bewick2012pipid}. This
analysis involves the species Xenopus, Silurana, Hymenochirus, Pipa and an
outgroup. Five species in total with a 4 taxa monophyly and a calibration on the
age of the parent of Pipa. There are $6\times3$ valid ranked topologies: $9$ of
those have $3$ internal nodes above the calibrated parent, $6$ has $2$ above and
$1$ below, and the remaining $3$ has $2$ below and one (the root) above.

The total density for $a$ internal nodes above and $b$ below by equation
\eqref{eqn:topmarginal} is
\begin{equation*}
f_{a,b}(h) = \lambda e^{-\lambda h} \frac{e^{-\lambda (a+1) h}}{(a+1)!} 
\frac{\left( 1 - e^{-\lambda h} \right)^b}{b!} ,
\end{equation*}
and so the marginal is:
\begin{align*}
\fM(h) = & \frac{5!}{18} \left( 9 f_{3,0}(h) + 6 f_{2,1}(h) + 3 f_{1,2}(h) \right)
\\
  = & \frac{5\lambda e^{-3\lambda h}}{6} \left( e^{-2\lambda h} - 4 e^{-\lambda
      h} + 6 \right) .
\end{align*}
   



\subsection{A four taxon tree with one calibration}

Following Heled and Drummond (2012) we consider the following four taxa tree in
which taxa \txA,\txB{} are constrained to be monophyletic and their most recent
common ancestor is calibrated with density $f_{\txAB}$.

There are 4 ranked topologies in this case, and the 2012 article gives the
marginal density for each. Here we wish to contrast the three priors using
concrete values: a birth rate of $\lambda = ^1/_2$ and a uniform calibration
prior ($f_{\txAB} = U[4,6]$). Table \ref{table:t1} summarises the results.

\if 11
\begin{center}
\begin{table*}[htp]
\centering
{\small{

\begin{tabular}{{@{} l l c c c @{}}} \toprule
  & & Multiplicative & Conditional & Restricted conditional \\
\midrule
Prior ``correction term'' & \txtr{((a,b),(c,d))} 
$T_{\txsb{\txc{}\txd{}}} < T_{\txsb{\txa{}\txb{}}}$
& -- 
&  $ 3\lambda e^{-3\lambda h_2} $ 
& $ 12 e^{-3 \lambda h_2} (1 - e^{-\lambda h_2}) $ \\
\cmidrule{2-5}
& \txtr{((a,b),(c,d))} $T_{\txsb{\txc{}\txd{}}} \ge T_{\txsb{\txa{}\txb{}}}$ 
& & & \\
& \txtr{(((\txa,b),c),d)} 
& -- 
& $ 3\lambda e^{-3\lambda h_3}$ 
& $4\lambda e^{-4\lambda h_3}$ \\

& \txtr{(((\txa,b),d),c)} & & & \\

\midrule
Marginal Topology probability & {\txtr{((a,b),(c,d))}} & $93\%$  & $94.2\%$ & $50\%$ \\
& {\txtr{(((a,b),c),d)}} & $3.5\%$ & $2.9\%$ & $25\%$ \\
& {\txtr{(((a,b),d),c)}} & $3.5\%$ & $2.9\%$ & $25\%$ \\
\midrule
Marginal calibration prior & & $\frac{3\lambda e^{-3\lambda x}}{e^{-12\lambda} -
  e^{-18\lambda}}$ & $\frac{1}{6-4}$ & $\frac{1}{6-4}$ \\
\bottomrule
\end{tabular}
\caption{An illustration of the difference between the restricted and full
  conditional prior using a 4 taxa example. The prior is a pure birth process
  with a birth rate of $\lambda = {^1/_2}$ and a uniform calibration density
  between 4 and 6 is applied to the monophyletic clade \txtr{(a,b)}. The uncorrected
  (multiplicative) prior is $\frac{1}{6-4} 4! \lambda^3 e^{-\lambda (2h_1 + h_2
    + h_3)}$, and the table gives the conditional prior ``correction terms'' for
  each ranked topology, together with the induced prior probability of each
  unranked topology and the marginal density for the calibrated clade.}
\label{table:t1} 
}}
\end{table*}
\end{center}
\fi

The table lists the ``correction term'' for each ranked topology, the marginal
probability for each unranked topology and the calibration marginal. As
expected, the full and restricted conditional preserve the calibration density,
while the marginal for the multiplicative prior is equal to the conditional
marginal ($3\lambda e^{-3\lambda x}$), bounded between 4 and 6. The marginal
topology probability illustrates the difference between the full and restricted
priors. The former is similar to the multiplicative prior, with a high
probability on the balanced tree. In the space of Yule trees with birth rate
$^1/_2$ and one internal node age between 4 and 6, the other age is far more
likely to be smaller that the first.
The latter, with equal weight for the two classes, matches the probabilities
of the Yule prior without calibration.

\subsection{Three Calibrations for Bombina}

A recent study using 13 complete genomes investigated the phylogenetic
relationships of the fire-bellied toads of the genus Bombina
\citep{pabijan2013complete}. The study contains many kinds of analysis, using
different calibration schemes. One BEAST analysis used three nested calibration
points, on 5 taxa, 7 taxa and the root. The results of running only the
multiplicative prior are shown in figure \ref{fig:bom} (a).

\ifinlf
\begin{figure}[htp!]
\centering
\includegraphics[width=16cm]{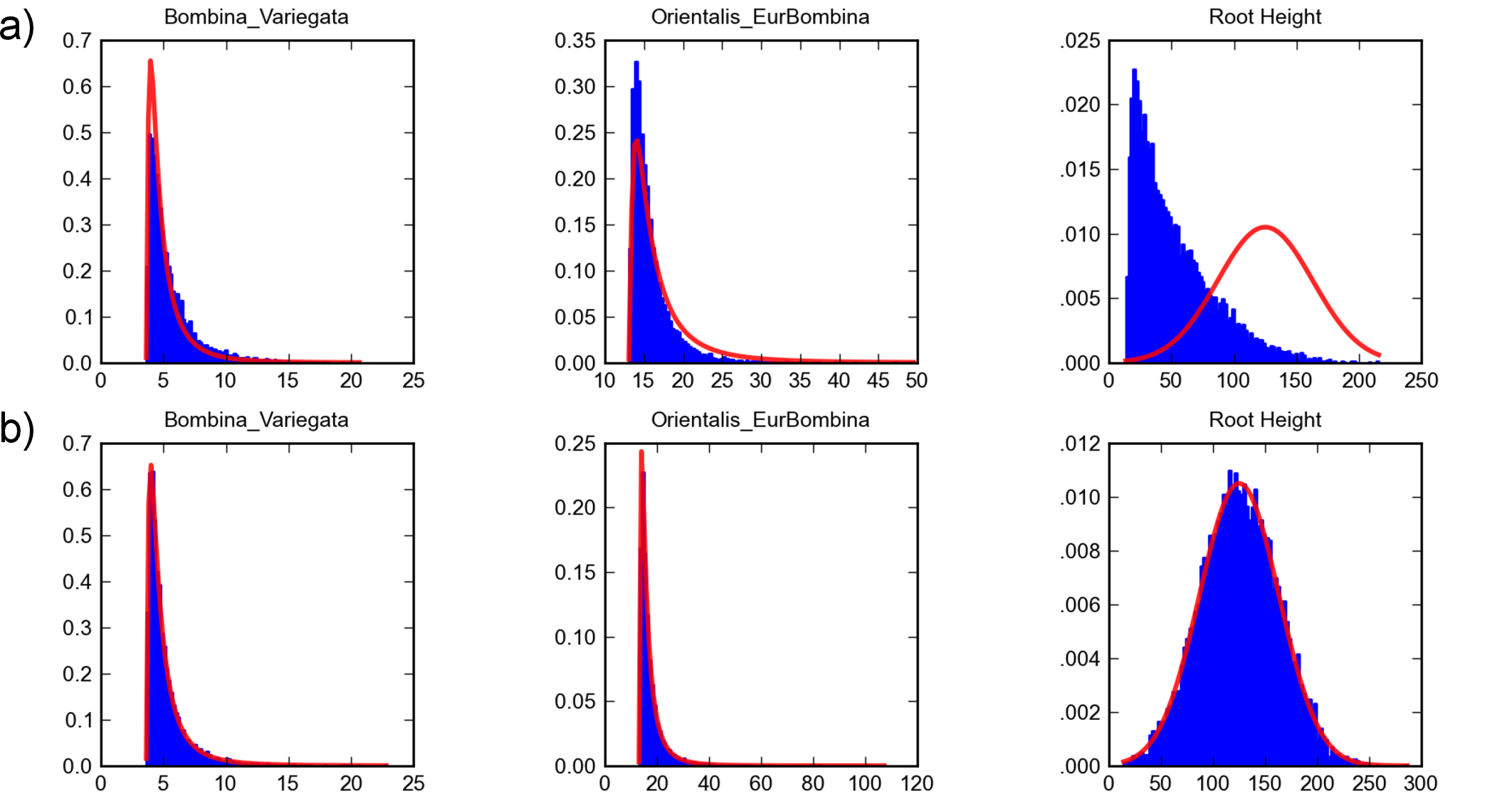}
\caption{Three calibration densities for the Bombina analysis, (a) under BEAST
multiplicative calibration prior and (b) under the conditional prior. The
specified densities are in red, and the blue show the density from a BEAST run.}
\label{fig:bom}
\end{figure}
\fi

While the marginal for the two clades deviate only slightly from the
calibration, the marginal for the root, with mean around 50, is way off from the
normal density calibration (N(125,36)). The marginals for the analysis
using the conditional prior described here
match the calibration densities as expected (Figure \ref{fig:bom} (b)).

\section{Conclusions}

We have presented a general approach to specifying a birth-death process tree
prior conditional on the heights of a set of calibrated nodes, in the context of
the joint inference of topology and divergence times.  We have described a few
special cases where this prior density has a closed form solution and we have
described a general, though computationally intensive, approach to numerical
calculation of this conditional density for any number of calibrated nodes. As a
result, an arbitrary marginal prior distribution can be precisely specified on
the calibrated nodes.

We have also described how the conditional birth-death tree prior naturally
induces a non-uniform distribution over ranked topologies.  If this effect is
unwanted, our approach can be modified to produce a uniform prior on ranked
topologies (therefore permitting any arbitrary distribution on ranked tree
topologies to be composed with the conditional birth-death prior on divergence
times).  This modification also renders a computationally efficient algorithm
for calculation of the prior density.

In order to compute the conditional birth-death prior, it is necessary to
compute the marginal density of the calibrated node heights averaged over all
consistent time-trees.  Although we have described some special cases where this
marginal prior density of the calibrated nodes can be efficiently computed, it
remains to be determined whether other cases have analytical closed-form
solutions.

Our implementation is available in BEAST2 (http://code.google.com/p/beast2/). We
regard the full conditional formulation as the correct approach, if one assumes
that the birth-death process prior is the appropriate prior for the phylogenetic
time-tree under estimation.  We therefore recommend the full conditional
formulation when computationally feasible (e.g. 2-3 calibrations and/or small
numbers of taxa).  The restricted formulation effectively removes influence of
the birth-death prior on the estimation of the ranked topology and is a good
alternative for analyses with large numbers of calibrations or taxa for which
computational considerations will preclude application of the full
conditional. Both of these approaches relieve the practitioner from running
their calibrated analysis in the absence of data in order to determine the
resulting marginal distributions (e.g. compare Fig \ref{fig:bom}a and Fig
\ref{fig:bom}b).

It is clear that development of calibrated tree priors for Bayesian phylogenetic
inference remains an area ripe for future development.  Obvious next steps would
include taking explicitly account of the different sources of uncertainty in
fossil ages and collection (uncertainty in geological dates, variation in fossil
preservation rates and paleontological discovery effort) and more sophisticated
means of dealing with the phylogenetic placement of fossil information
(uncertain placement of fossils based on morphological characters of fossils
and/or tree process prior assumptions).  All of these factors are currently
subsumed into whatever marginal distribution is specified on the set of
calibrated nodes.  In the mean time the work presented here derives new results
for multiple-calibration tree priors and in doing so illustrates some of the
subtle choices open to the practitioner when calibrating birth-death tree
priors.

\section{Authors contributions}
AJD and JH designed the study. JH designed the algorithm for computation of full
conditional formulation. AJD proposed the solution for the restricted
conditional formulation. JH produced all derivations and performed all data
analyses. AJD and JH produced the figures and wrote the manuscript.

\section{Acknowledgements}

The authors wish to thank Mike Steel for helpful discussion and comments. JH and
AJD were supported by a Rutherford Discovery Fellowship from the Royal Society
of New Zealand awarded to AJD.

\bibliographystyle{plainnat}

\begin{thebibliography}{24}
\providecommand{\natexlab}[1]{#1}
\providecommand{\url}[1]{\texttt{#1}}
\expandafter\ifx\csname urlstyle\endcsname\relax
  \providecommand{\doi}[1]{doi: #1}\else
  \providecommand{\doi}{doi: \begingroup \urlstyle{rm}\Url}\fi

\bibitem[Bewick et~al.(2012)Bewick, Chain, Heled, and Evans]{bewick2012pipid}
Adam~J Bewick, Fr{\'e}d{\'e}ric~JJ Chain, Joseph Heled, and Ben~J Evans.
\newblock The pipid root.
\newblock \emph{Systematic Biology}, 61\penalty0 (6):\penalty0 913--926, 2012.

\bibitem[Drummond and Rambaut(2007)]{BEAST}
A.J Drummond and A~Rambaut.
\newblock Beast: Bayesian evolutionary analysis by sampling trees.
\newblock \emph{BMC Evol Biol}, 7:\penalty0 214, 2007.
\newblock ISSN 1471-2148 (Electronic).
\newblock \doi{10.1186/1471-2148-7-214}.

\bibitem[Drummond et~al.(2006)Drummond, Ho, Phillips, and
  Rambaut]{drummond2006relaxed}
AJ~Drummond, SYW Ho, MJ~Phillips, and A.~Rambaut.
\newblock {Relaxed phylogenetics and dating with confidence}.
\newblock \emph{PLoS Biol}, 4\penalty0 (5):\penalty0 e88, 2006.

\bibitem[Felsenstein(1981)]{felsenstein1981evolutionary}
J.~Felsenstein.
\newblock {Evolutionary trees from DNA sequences: a maximum likelihood
  approach}.
\newblock \emph{Journal of molecular evolution}, 17\penalty0 (6):\penalty0
  368--376, 1981.

\bibitem[Gernhard(2008)]{gernhard2008conditioned}
T.~Gernhard.
\newblock {The conditioned reconstructed process}.
\newblock \emph{Journal of Theoretical Biology}, 253\penalty0 (4):\penalty0
  769--778, 2008.

\bibitem[Heled and Drummond(2012)]{heledDrummond2012}
Joseph Heled and Alexei~J. Drummond.
\newblock Calibrated tree priors for relaxed phylogenetics and divergence time
  estimation.
\newblock \emph{Systematic Biology}, 61\penalty0 (1):\penalty0 138--149, 2012.

\bibitem[H{\"o}hna et~al.(2011)H{\"o}hna, Stadler, Ronquist, and
  Britton]{Hohna:2011kx}
Sebastian H{\"o}hna, Tanja Stadler, Fredrik Ronquist, and Tom Britton.
\newblock Inferring speciation and extinction rates under different sampling
  schemes.
\newblock \emph{Mol Biol Evol}, 28\penalty0 (9):\penalty0 2577--89, Sep 2011.
\newblock \doi{10.1093/molbev/msr095}.

\bibitem[Huelsenbeck and Ronquist(2001)]{huelsenbeck2001mbi}
J.P. Huelsenbeck and F.~Ronquist.
\newblock {MRBAYES: Bayesian inference of phylogenetic trees}, 2001.

\bibitem[Kendall(1948)]{kendall1948generalized}
David~G Kendall.
\newblock On the generalized" birth-and-death" process.
\newblock \emph{The annals of mathematical statistics}, pages 1--15, 1948.

\bibitem[Kishino et~al.(2001)Kishino, Thorne, and Bruno]{kishino:2001:pe}
Hirohisa Kishino, Jeffrey~L. Thorne, and William~J. Bruno.
\newblock Performance of a divergence time estimation method under a
  probabilistic model of rate evolution.
\newblock \emph{Molecular Biology and Evolution}, 18\penalty0 (3):\penalty0
  352--361, 2001.
\newblock URL
  \url{http://mbe.oxfordjournals.org/cgi/content/abstract/18/3/352}.

\bibitem[Nee et~al.(1994{\natexlab{a}})Nee, Holmes, May, and Harvey]{Nee1994b}
S~Nee, E~C Holmes, R~M May, and P~H Harvey.
\newblock Extinction rates can be estimated from molecular phylogenies.
\newblock \emph{Philos Trans R Soc Lond B Biol Sci}, 344\penalty0
  (1307):\penalty0 77--82, Apr 1994{\natexlab{a}}.
\newblock \doi{10.1098/rstb.1994.0054}.

\bibitem[Nee et~al.(1994{\natexlab{b}})Nee, May, and Harvey]{Nee:1994fk}
S~Nee, R~M May, and P~H Harvey.
\newblock The reconstructed evolutionary process.
\newblock \emph{Philos Trans R Soc Lond B Biol Sci}, 344\penalty0
  (1309):\penalty0 305--11, May 1994{\natexlab{b}}.
\newblock \doi{10.1098/rstb.1994.0068}.

\bibitem[Pabijan et~al.(2013)Pabijan, Wandycz, Hofman, Wecek, Piwczy{\'n}ski,
  and Szymura]{pabijan2013complete}
Maciej Pabijan, Anna Wandycz, Sebastian Hofman, Karolina Wecek, Marcin
  Piwczy{\'n}ski, and Jacek~M Szymura.
\newblock Complete mitochondrial genomes resolve phylogenetic relationships
  within< i> bombina</i>(anura: Bombinatoridae).
\newblock \emph{Molecular phylogenetics and evolution}, 2013.

\bibitem[Rannala and Yang(1996)]{Rannala:1996uq}
B~Rannala and Z~Yang.
\newblock Probability distribution of molecular evolutionary trees: a new
  method of phylogenetic inference.
\newblock \emph{J Mol Evol}, 43\penalty0 (3):\penalty0 304--11, Sep 1996.

\bibitem[Rannala and Yang(2007)]{Rannala2007}
Bruce Rannala and Ziheng Yang.
\newblock Inferring speciation times under an episodic molecular clock.
\newblock \emph{Systematic Biology}, 56\penalty0 (3):\penalty0 453--466, 2007.

\bibitem[Sprugnoli(2006)]{sprugnoli2006introduction}
Renzo Sprugnoli.
\newblock An introduction to mathematical methods in combinatorics.
\newblock \emph{Dipartimento di Sistemi e Informatica Viale Morgagni}, 2006.

\bibitem[Stadler(2009{\natexlab{a}})]{Stadler:2009fk}
Tanja Stadler.
\newblock On incomplete sampling under birth-death models and connections to
  the sampling-based coalescent.
\newblock \emph{J Theor Biol}, 261\penalty0 (1):\penalty0 58--66, Nov
  2009{\natexlab{a}}.
\newblock \doi{10.1016/j.jtbi.2009.07.018}.

\bibitem[Stadler(2009{\natexlab{b}})]{stadler2009incomplete}
Tanja Stadler.
\newblock On incomplete sampling under birth--death models and connections to
  the sampling-based coalescent.
\newblock \emph{Journal of Theoretical Biology}, 261\penalty0 (1):\penalty0
  58--66, 2009{\natexlab{b}}.

\bibitem[Steel and Mooers(2010)]{steel2010expected}
Mike Steel and Arne Mooers.
\newblock The expected length of pendant and interior edges of a yule tree.
\newblock \emph{Applied Mathematics Letters}, 23:\penalty0 1315--1319, 2010.
\newblock \doi{DOI 10.1016/j.aml.2010.06.021}.

\bibitem[Thorne and Kishino(2002)]{Thorne:2002ff}
Jeffrey~L Thorne and Hirohisa Kishino.
\newblock Divergence time and evolutionary rate estimation with multilocus
  data.
\newblock \emph{Syst Biol}, 51\penalty0 (5):\penalty0 689--702, 2002.
\newblock ISSN 1063-5157 (Print).
\newblock \doi{10.1080/10635150290102456}.

\bibitem[Thorne et~al.(1998)Thorne, Kishino, and Painter]{Thorne1998}
J.L. Thorne, H.~Kishino, and I.S. Painter.
\newblock Estimating the rate of evolution of the rate of molecular evolution.
\newblock \emph{Molecular Biology and Evolution}, 15:\penalty0 1647--1657,
  1998.

\bibitem[Yang and Rannala(1997)]{Yang:1997uq}
Z~Yang and B~Rannala.
\newblock Bayesian phylogenetic inference using dna sequences: a markov chain
  monte carlo method.
\newblock \emph{Mol Biol Evol}, 14\penalty0 (7):\penalty0 717--24, Jul 1997.

\bibitem[Yang and Rannala(2006)]{yang2006bayesian}
Ziheng Yang and Bruce Rannala.
\newblock Bayesian estimation of species divergence times under a molecular
  clock using multiple fossil calibrations with soft bounds.
\newblock \emph{Molecular Biology and Evolution}, 23\penalty0 (1):\penalty0
  212--226, 2006.

\bibitem[Yule(1924)]{Yule1924}
G.U. Yule.
\newblock A mathematical theory of evolution based on the conclusions of
  dr.~j.c.~willis.
\newblock \emph{Philosophical Transactions of the Royal Society of London,
  Series B}, 213:\penalty0 21--87, 1924.

\end{thebibliography}

\appendix

\section{Two Monophyletic Nested Clades for the Yule Prior}
\label{app:twoclades}

$\NR_n^k$ is the number of ranked ways $n$ lineages can coalesce to $k$
\begin{equation}\label{eq:rnk}
\begin{split}
\NR_n =  & \prod_{i=2}^n {i \choose {2}} = \frac{n! (n - 1)!}{2^{n - 1}} \\
\NR_n^k = & \prod_{i=k+1}^{n} {i \choose {2}} = \frac{\NR_n}{\NR_k} =
 2^{-(n - k)} \frac{n{(n - 1)!}^2}{k{(k-1)!}^2}
\end{split}
\end{equation}

\subsection{Root and Clade}

For the marginal of a monophyletic clade of $n$ taxa and the root in a $n+m$
taxa tree we partition $\SRT_\C$ so that $\SRT_\C^k$ contain all topologies with
$k+1$ surviving lineages at time $h$ (Figure \ref{fig:rootAndClade}).

The size of each subset is
\begin{equation}
| \SRT_\C^k | = \binom{n-2+l-k}{n-2} \NR_n \NR_l^k \NR_{k+1} 
\end{equation}

and from (\cite{heledDrummond2012} appendix C, equation (12)) we have
\begin{equation}\label{eq:rtln}
| \SRT_\C | = \binom{n+l}{l-1} \NR_l \NR_n . 
\end{equation}

Plugging those counts into equation \eqref{eqn:marginal-split} we get
{\small{
\begin{equation}
\begin{split}
  \frac{1}{| \SRT_\C |} & \sum_{k=1}^l \left [ | \SRT_\C^k | (n+l)!
\lambda  e^{-\lambda h} 
\lambda  e^{-2 \lambda h_0} \right . \\
 & \left . \frac{(1-e^{-\lambda h})^{n-2+l-k}}{(n-2+l-k)!}
\frac{(e^{\lambda h} - e^{\lambda h_0})^{k-1}}{(k-1)!} \right ] = \\
  (n - 1) & n(n + 1) \sum_{k=1}^l  \left [ \binom{k + 1}{2} \binom{l-1}{k-1} 
\lambda  e^{-\lambda h} 
\lambda  e^{-2 \lambda h_0} \right . \\
& \left . (1-e^{-\lambda h})^{n-2+l-k}
(e^{\lambda h} - e^{\lambda h_0})^{k-1} \right ]  = \\
(n - 1)& n(n + 1) \lambda^2  e^{-\lambda( h + 2 h_0)} (1-e^{-\lambda h})^{n-2} \\
 & \sum_{k=1}^l \binom{k + 1}{2} \binom{l-1}{k-1}  
(1-e^{-\lambda h})^{l-k}
(e^{\lambda h} - e^{\lambda h_0})^{k-1} ,
\end{split}
\end{equation}
}}

which simplifies to equation \eqref{eqn:rootandclade}, because without the
$\binom{k + 1}{2}$, the sum is the binomial expansion of $(u+v)^{l-1}$, and
with the combinatorial identity
\begin{equation}\label{eq:comb1}
\sum_{k=0}^n (k)_{m} \binom{n}{k} u^k v^{n-k} = (n)_{m} u^{m} (u+v)^{n-m} ,
\end{equation}
we can simplify such sums where the terms are multiplied by any simple
polynomial in $k$.

$(x)_n$ is the Pochhammer symbol, the falling factorial. Here $\binom{k + 1}{2}
= {^1/_2} (k)_{2} + (k)_{1}$.

\subsection{Two Nested Clades}

When the top clade is not the root we need to handle 3 levels. Let the number of
surviving lineages at $h_2$ be $m_1$ and $l_1$, and $l_2$ at $h_1$ (Figure
\ref{fig:twonested}). We partition $\SRT_\C$ according to $m_1$, $l_1$ and
$l_2$. that is topologies with the equal values are in the same class.

The number of internal nodes at the three levels is
\begin{equation*}
\begin{split}
k_0 = &  n+m+l - (m_1+l_1+2) \\
k_1 = &  m_1+l_1 - (1+l_2) \\
k_2 = &  l_2+1 - 1 .
\end{split}
\end{equation*}

The size of each subset is 
{
\small{
\begin{equation}
\begin{split}
| \SRT_\C^{m_1,l_1,l_2} | = & \NR_n \NR_m^{m_1} \NR_l^{l_1}
\frac{k_0!}{(n-2)!(m-m_1)!(l-l_1)!} \\ 
& \NR_{m_1+1} \NR_{l_1}^{l_2} \binom{k_1}{m_1-1} \\
& \NR_{l_2 + 1} .
\end{split}
\end{equation}
}}
Each of the 3 lines above gives the contribution of one level. The total number
of topologies is
\begin{equation}
| \SRT_{\C} | = \binom{n+m}{m-1} \NR_m \NR_n  \binom{n+m+l}{l-1}  \NR_l .
\end{equation}

This can be obtained either from summing over all $\SRT_\C^{m_1,l_1,l_2}$ terms
or more simply by applying equation \eqref{eq:rtln} twice, since the internal
clade $N$ does not interact with the free global lineages. Again pluggin those
counts into equation \eqref{eqn:topmarginal} we get
\begin{equation}
\begin{split}
f_{m_1,l_1,l_2}(h_1,h_2) = & (n+m+l)! \lambda^2 e^{-\lambda(h_2 + h_1)} 
 \frac{e^{-(k_2+1) \lambda h_1}}{(k_2+1)!} \\
& \frac{(1 - e^{-\lambda h_2}){^{k_0}}}{k_0!}  
 \frac{(e^{-\lambda h_2} - e^{-\lambda h_1})^{k_1}}{k_1!}  .
\end{split}
\end{equation}

And finally
\begin{equation}
\begin{split}
f(h_1,h_2) = & {| \SRT_{\C}|}^{-1}
 \sum_{m_1 = 1}^m \sum_{l_1=1}^l \sum_{l_2=1}^{l_1} | \SRT_\C^{m_1,l_1,l_2} |
 f_{m_1,l_1,l_2} .
\end{split}
\end{equation}

The rest is tedious manipulations similar to those in the root and clade case above.

\section{Integral Identity used in Obtaining the Yule Marginal}
\label{app:intkabove}

\begin{equation}
\label{eqn:intkabove}
\int_h^{\infty} n \lambda e^{- n\lambda x} (e^{-\lambda h} - e^{-\lambda x})^m
\,\D x
= \binom{m+n}{n}^{-1} e^{-(m+n) \lambda h}
\end{equation}

Proof:
{\small
\begin{equation*}
\begin{split}
\int_h^{\infty} n \lambda e^{- n\lambda x} 
(e^{-\lambda h} - e^{-\lambda x})^m \,\D x & \quad = \int_h^{\infty}
n \lambda e^{- n\lambda x} e^{-m \lambda h} (1 - e^{-\lambda (x-h)})^m \,\D x \\
& \quad = \int_h^{\infty} n \lambda e^{- n\lambda x} e^{-m \lambda h} 
\sum_{k=0}^m (-1)^k \binom{m}{k} e^{-k \lambda (x-h)} \,\D x \\
& \quad =  n \lambda e^{-m \lambda h} 
\sum_{k=0}^m (-1)^k \binom{m}{k} e^{\lambda k h} \int_h^{\infty} e^{- (k+n)\lambda x} \,\D x  \\
& \quad =  n \lambda e^{-m \lambda h} 
\sum_{k=0}^m (-1)^k \binom{m}{k} e^{\lambda k h} \frac{e^{- (k+n)\lambda h}}{(k+n)\lambda}   \\
& \quad =  e^{-(m+n) \lambda h} 
\sum_{k=0}^m (-1)^k \binom{m}{k} \frac{n}{(k+n)}  \quad \text{Using \eqref{eqn:comb01}}\\
& \quad =  e^{-(m+n) \lambda h} \binom{m+n}{n}^{-1} 
\end{split}
\end{equation*}
}

The last step used the well known combinatorial identity (for example,
\cite{sprugnoli2006introduction}, page 74)
\begin{equation}
\label{eqn:comb01}
\sum_{k=0}^m (-1)^k \binom{m}{k} \frac{n}{n+k} = \binom{m+n}{n}^{-1} .
\end{equation}

\section{The Birth-Death Prior Marginal}

For convenience, let $z = x_{i_1}$ be the age of the last calibration point, and
$\hat{c} = c_1 - 1$, the number of lineages between the root and the last
calibration point (excluding the root).

\begin{equation}
\begin{aligned}
P_0(z) = \int_z^{\infty} \int_z^{x_1} \cdots \int_z^{x_{\hat{c}}} 
\left [ q_1(x_1) \prod_{k=1}^{\hat{c}+1}  p_1(x_k) \right ] d x & = \\
\int_z^\infty \left [ q_1(x_1) p_1(x_1)  \frac{\left ( P_1(x_1) - P_1(z) \right )
    ^{\hat{c}}}{\hat{c} !} \right ] d x_1 & = \\
\frac{1}{\hat{c} !} \int_z^\infty  q_1(x_1) p_1(x_1) 
  \sum_{j=0}^{\hat{c}} \binom{\hat{c}}{j} {P_1(x_1)}^j
        (-P_1(z))^{\hat{c} - j} d x_1 & = \\
\frac{1}{\hat{c} !} \sum_{j=0}^{\hat{c}} (-P_1(z))^{\hat{c} - j}  \binom{\hat{c}}{j}
\int_{z}^{\infty} q_1(x_1) p_1(x_1){P_1(x_1)}^j d x_1 .
\end{aligned}
\label{eqn:rootdens}
\end{equation}

The following solution for the integral can be verified by taking the derivative
of the right hand side
\begin{align*}
\int q_1(x_1) p_1(x_1){P_1(x_1)}^j d x_1 & = 
\frac{1}{(j+1)(j+2)}  \left (\frac{\mu - \lambda}{\mu'} \right )^{j+2}
 \frac{\lambda' - (j+2) \mu' e^{-(\lambda - \mu) x_1}} 
{ (\lambda' - \mu' e^{ -(\lambda - \mu) x_1})^{j+2}} \\
& = \frac{\mu'^{-(j+2)}}{(j+1)(j+2)} (\lambda' - (j+2) \mu' e^{-(\lambda - \mu) x_1}) q(x_1)^{j+2}.
\end{align*}

Substituting in Eq. \eqref{eqn:rootdens} and simplifying gives

\begin{equation}
\begin{aligned}
P_0(z,x_1) =
\frac{1}{\hat{c} !} \sum_{j=0}^{\hat{c}} (-P_1(z))^{\hat{c} - j}  \binom{\hat{c}}{j} 
\frac{\mu'^{-(j+2)}}{(j+1)(j+2)} 
 (\lambda' - (j+2) \mu' e^{-(\lambda - \mu) x_1}) q(x_1)^{j+2} & = \\
\frac{1}{\hat{c}!} \sum_{j=0}^{\hat{c}}
 \left( \frac{-q(z)}{\mu'} \right)^{\hat{c}-j} \binom{\hat{c}}{j} 
\frac{\mu'^{-(j+2)}}{(j+1)(j+2)} (\lambda' - (j+2) \mu' e^{(\lambda - \mu) x_1}) q(x_1)^{j+2} & = \\
\frac{\mu'^{-(\hat{c}+2)}}{(\hat{c}+2) !} 
\sum_{j=0}^{\hat{c}} \binom{\hat{c}+2}{j+2} (-q(z))^{\hat{c}-j}
 (\lambda' - (j+2) \mu' e^{(\lambda - \mu) x_1}) q(x_1)^{j+2} & = \\
\frac{\mu'^{-(\hat{c}+2)}}{(\hat{c}+2) !} 
\left[ \sum_{j=0}^{\hat{c}}  \lambda' \binom{\hat{c}+2}{j+2} (-q(z))^{\hat{c}-j} q(x_1)^{j+2} 
- \sum_{j=0}^{\hat{c}} -(j+2) \binom{\hat{c}+2}{j+2} \mu' e^{(\lambda - \mu)
} (-q(z))^{\hat{c}-j} q(x_1)^{j+2} \right] & = \\
\frac{\mu'^{-(\hat{c}+2)}}{(\hat{c}+2) !} \lambda' 
\left( (q(x_1) + q(z))^{\hat{c}+2} - (-q(z))^{\hat{c}+2} - (\hat{c}+2) q(x_1)
(-q(z))^{\hat{c}+1} \right) \\
+   \left( (\hat{c}+2) q(x_1) ( (-q(z))^{\hat{c}+1} - (q(x_1) +
  q(z))^{\hat{c}+1} ) \right) .
\end{aligned}
\end{equation}

The last step uses Equation \eqref{eq:comb1} for the second sum. 
Now, after canceling terms and simplifying we are left with 
\begin{equation}
\begin{aligned}
P_0(z) = P_0(z,x_1) \Big|_{x_1=z}^{\infty} = \frac{\lambda'^{-(k+1)}}{(k+2)!} q_1(z)^{k+2} .
\end{aligned}
\end{equation}

\ifinlf
\else
\section{Figures}
  \subsection*{Figure 1 - Notation}

For the tree above we have: $n=7$ taxa, the ranked topology
$\RT=$\txtr{(((a,(b,c):1):3,d):5,((e,f):2,g):4)} in NEWICK format, with internal
nodes marked by rank, $\C= (\{\txtr{e,f,g}\}, \{\txtr{b,c}\},\{\txtr{a,b,c,d}\})$,
$\nc=3$ calibrated nodes marked in red, and $\TR{}(\RT) = (3,6,2)$, $\RA{}(\RT)
= (2,1,3)$, $\AR{}(\RT) = (2,3,6)$ and $h_\RT = (h_3,h_6,h_2)$.  

\captionsetup[figure]{labelformat=empty}
\begin{figure}
\caption{\label{fig:notation}}
\end{figure}

  \subsection*{Figure 2 - Counting Ranked Topologies}

To count the number of ranked topologies for the tree above, we multiply the
counts in the 3 levels. In the lowest level we have 3 lineages reducing to 1
(root of lowest calibration), 5 lineages reducing to 3 and 2 free lineages not
reducing. So, the total number of topologies is

$\NR_3^1 \NR_5^2 \NR_2^2 \binom{2+5+2 - (1+3+2)}{1,2,0} = 3 \times (10\times 6)
\times 1 \times \frac{3!}{1!2!0!} = 540$

Note that in the binomial we use one less lineage (2 instead of 3) for the
calibrated clade, since its position as root is fixed. 

In the second level we have 3 lineages reducing to  1, and 3 free lineages
reducing to 2, giving

$\NR_3^1 \NR_3^2 \binom{2+3 - (1+2)}{1,1} = 3 \times 3 \times \frac{2!}{1!1!} =
9$

and in the last level 3 lineages to 1 in 3 ways. So, the total number is $540
\times 9 \times 3 = 12960$.

\begin{figure}
\caption{\label{fig:counting}} 
\end{figure}

  \subsection*{Figure 3 - Nested Calibrations Iterator}

The calibrated clades are N (green), M (blue) and L (red), with respectively
$n+2$, $n+2+m+1$ and $l+2$ taxa. In addition we have $r$ ``free'' taxa in
orange, which can coalesce between themselves and with the roots of M and
L. There are 4 levels associated with the 3 calibration nodes, separated by the
root ages of the calibrated clades.

\begin{figure}
\caption{\label{fig:figxx}}
\end{figure}

  \subsection*{Figure 4 - Parent of Monophyletic Clade}

Parent of monophyletic clade of size $n$.

\begin{figure}
\caption{\label{fig:parentOfClade} }
\end{figure}

  \subsection*{Figure 5 - Root and Monophyletic Clade}

  Monophyletic clade of size $n$ and root with $n+m$ taxa.

\begin{figure}
\caption{\label{fig:rootAndClade} }
\end{figure}

  \subsection*{Figure 6 - Two Nested Monophyletic Clades}

Two nested monophyletic clades of size $n$ and $n+m$ taxa in a $n + m
  + l$ taxa tree ($l > 0$).

\begin{figure}
\caption{\label{fig:twonested} }
\end{figure}

  \subsection*{Figure 7 - Three Calibrations for Bombina}

Three calibration densities for the Bombina analysis, (a) under BEAST
multiplicative calibration prior and (b) under the conditional prior. The
specified densities are in red, and the blue show the density from a BEAST run.

\begin{figure}
\caption{\label{fig:bom} }
\end{figure}

\fi

\end{document}